\documentclass{pasj01}
\Received{2015/11/11}
\Accepted{2016/01/17}
\Published{$\langle$publication date$\rangle$}
\SetRunningHead{Astronomical Society of Japan}{Usage of \texttt{pasj00.cls}}

\begin{document}
\title{
Suzaku Follow-up of Heavily Obscured Active Galactic Nuclei
Detected in Swift/BAT Survey: NGC 1106, UGC 03752, and NGC 2788A}
\author{
Atsushi Tanimoto	\altaffilmark{1}, 
Yoshihiro Ueda		\altaffilmark{1}, 
Taiki Kawamuro		\altaffilmark{1}, 
Claudio Ricci		\altaffilmark{1,2}}
\altaffiltext{}{
\altaffilmark{1}Department of Astronomy, Kyoto University, Kyoto 606-8502\\
\altaffilmark{2}Instituto de Astrofisica, Pontificia Universidad Catolica de Chile, Casilla 306, Santiago 22, Chile}
\email{tanimoto@kusastro.kyoto-u.ac.jp}
\maketitle

\begin{abstract}
We present the broadband (0.5--100 keV) spectra of three heavily obscured Active Galactic Nuclei (AGNs), NGC 1106, UGC 03752, and NGC 2788A, observed with Suzaku and Swift/Burst Alert Telescope (BAT). The targets are selected from the Swift/BAT 70-month catalog on the basis of high hardness ratio between above and below 10 keV, and their X-ray spectra are reported here for the first time. We apply three models, a conventional model utilizing an analytic reflection code and two Monte-Carlo based torus models with a doughnut-like geometry (MYTorus; \cite{Murphy09}) and with a nearly spherical geometry (Ikeda torus; \cite{Ikeda09}). The three models can successfully reproduce the spectra, while the Ikeda torus model gives better description than the MYTorus model in all targets. We identify that NGC 1106 and NGC 2788A as Compton-thick AGNs. We point out that the common presence of unabsorbed reflection components below 7.1 keV in obscured AGNs, as observed from UGC 03752, is evidence for clumpy tori. This implies that detailed studies utilizing clumpy torus models are required to reach correct interpretation of the X-ray spectra of AGNs. 
\end{abstract}

\section{Introduction}
\begin{table*}
\caption{List of Target}
\begin{center}
\begin{tabular}{lcccc}
\hline
\multicolumn{1}{c}{} & & J0250.7+4142 & J0714.2+3518 & J0902.7-6816				\\
\hline
(1)	& Optical ID
& NGC 1106				& UGC 03752				& NGC 2788A			\\
(1)	& Classification
& Seyfert 2				& Seyfert 2				& Galaxy				\\
(1)	& Redshift
& 0.0144					& 0.0157					& 0.0133				\\
(2)	& $N_{\mathrm{H}}^{\mathrm{Gal}}$ [10$^{22}$ cm$^{-2}$]
& 0.0735					& 0.0930 					& 0.0152				\\
(3)	& Start Time	[UT]
& 2014 August 3 01:18		& 2015 April 5 23:51			& 2015 May 15 01:08		\\
(3)	& End Time	[UT]
& 2014 August 4 21:09		& 2015 April 6 21:42			& 2015 May 16 05:21		\\
(3)	& Exposure	[ks]
& 67.0					& 67.4					& 23.6				\\
\hline
\multicolumn{1}{@{}l@{}}{\hbox to 0pt{\parbox{160mm}
{\footnotesize
\textbf{Notes.}\\
(1)Taken from the NASA/IPAC Extragalactic Database.	\\
(2)The hydrogen column density of Galactic absorption \citep{Kalberla05}.\\
(3)Based on the good time interval for XIS-0.\\
}\hss}}
\end{tabular}
\end{center}
\end{table*}

Heavily obscured Active Galactic Nuclei (AGNs) are key objects to understand the origin of the cosmic X-ray background (e.g., \cite{Ueda03, Gilli07}) and the ``co-evolution'' of Supermassive Black Holes (SMBHs) and galaxies \citep{Kormendy13}, which may have rapidly grown during the obscuration phase after major mergers (e.g., \cite{Sanders88, Hopkins06}). A generally accepted picture of AGN structure is that a central SMBH and an accretion disk are surrounded by obscuring matter on parsec scales consisting of gas and dust, called ``torus'' \citep{Antonucci85, Antonucci93}. While a torus is believed to play a crucial role for gas feeding onto a SMBH (e.g., \cite{Kawakatu08, Wada15}), its origin has been poorly understood. For instance, it is also proposed that a torus may be a remnant of outflow launched from the accretion disk \citep{Elitzur06}, or a tilted disk \citep{Lawrence10}. It is thus very important to constrain basic properties (e.g., geometry) of AGN tori from observations.

X-ray observations are a powerful tool to study the inner structure of an AGN, because X-rays can trace all material including gas and dust. A torus significantly affects the observed X-ray spectra by photoelectric absorption of the direct component when they block the light-of-sight (i.e., in type 2 AGNs) \citep{Awaki91}. In addition, reflected components from the torus with fluorescence iron-K lines are expected, which are more easily observable when the direct emission is absorbed. Hence, observations of heavily obscured AGNs, in particular Compton-thick AGNs, whose hydrogen column densities along the line-of-sight exceed $\log N_{\mathrm{H}} \geq 24 \ \mathrm{cm^{-2}}$, are very useful to investigate the torus geometry.

Surveys in hard X-rays above 10 keV provide one of the least biased sample of AGNs, thanks to their high penetrating power against obscuration as far as they are not heavily Compton-thick ($\log N_{\mathrm{H}} \geq 24.5 \ \mathrm{cm^{-2}}$). X-ray follow-up observations covering lower energies of all-sky hard X-ray surveys with Swift/Burst Alert Telescope (BAT) and INTEGRAL have produced many important results, such as column density distribution of local AGNs (e.g., \cite{Markwardt05, Beckmann06, Winter09a, Burlon11, Vasudevan13, Ueda14, Ricci15}) and simultaneous broadband spectra (e.g., \cite{Ueda07, Eguchi09, Winter09b, Tazaki13, Kawamuro13}). Because even hard X-ray fluxes are attenuated in the case of very large absorption, however, the observed number of Compton-thick AGNs was quite limited in shallow survey data, like the Swift/BAT 9-month catalog \citep{Tueller08}. In
fact, up to present, X-ray spectra of only a handful new Compton-thick AGNs selected in the hard X-ray surveys have bee n investigated in detail (e.g., \cite{Gandhi15}).

C.\ Ricci et al.\ (2016, in preparation) are working on a systematic X-ray spectroscopic survey of the Swift/BAT 70-month catalog, one of the deepest all-sky hard X-ray survey currently available \citep{Baumgartner13}, using data mainly from Swift/X-Ray Telescope (XRT) below 10 keV. One of main goals of this project is to establish the properties of Compton-thick AGNs in the local universe by significantly increasing their sample size. Because typical exposures of the Swift/XRT observations are short, however, longer follow-up observations using large area X-ray telescopes such as Suzaku and XMM-Newton are crucial to robustly identify new Compton-thick AGNs through accurate determination of the column densities and to understand the nature of their broadband spectra.

In this paper, we study the spectra of three heavily obscured AGNs, NGC 1106, UGC 03752, and NGC 2788A, observed with Suzaku and Swift/BAT. They are selected from the Swift/BAT 70-month catalog on the basis of the large flux ratio between above and below 10 keV. Their broadband X-ray data are reported here for the first time, and we identify NGC 1106 and NGC 2788A as new Compton-thick AGNs, and UGC 03752 as an Compton-thin heavily absorbed AGN. We apply ``torus'' models based on Monte Carlo simulations by \citet{Murphy09} and \citet{Ikeda09} to the Suzaku and Swift/BAT spectra, and discuss the degeneracy and limitation of these models and future prospects.

The organization of this paper is as follows. Section 2 describes the observations of the targets. In section 3, we present the results of spectral analysis. In Section 4, implications from our results are discussed. Section~5 gives the conclusion. The luminosities are calculated from the observed redshifts with the cosmological parameters $(H_{0}, \Omega_{\mathrm{m}}, \Omega_{\lambda})$ = (70 km s$^{-1}$ Mpc$^{-1}$, 0.3, 0.7). The solar abundances by \citet{Anders89} are assumed in all cases. In modeling photoelectric absorption, we adopt the cross section by \citet{Balucinska92}. The errors attached to spectral parameters correspond to 90\% confidence limits for a single parameter of interest.

\section{Targets and Observations}
We selected Compton-thick AGN candidates from the Swift/BAT 70-month catalog with the following criteria: (1) very weak soft X-ray flux (2--10 keV) relative to the hard X-ray flux (15--100 keV) by $F_{2-10}/F_{15-100} < 0.03$, which is a good indicator for heavily obscured AGNs \citep{Winter09a}, (2) indication of a strong narrow iron-K line in the Swift/XRT spectrum of a short exposure (typically $\sim$ 10 ksec), and (3) optical classification as Seyfert 2 or galaxy. Our targets, Swift J0250.7+4142 (NGC 1106), Swift J0714.2+3518 (UGC 03752), and Swift J0902.7-6816 (NGC 2788A), correspond to brightest ones in the 15--100 keV band among those candidates whose spectra below 10 keV have not been reported previously. The basic information of the targets is summarized in Table~1.
 
We observed these targets in 2014 and 2015 with Suzaku \citep{Mitsuda07}, the fifth observatory in a series of Japanese X-ray astronomy satellites. Table~1 gives the observation log. Suzaku carries four X-ray CCD cameras called the X-ray Imaging Spectrometers (XIS-0, XIS-1, XIS-2, and XIS-3) covering the 0.2--12 keV as the focal plane detectors of four X-ray telescopes, and a non-imaging instrument called the Hard X-ray Detector (HXD), consisting of Si PIN photodiodes and Gadolinium Silicon Oxide scintillation counters, which covers the 10--70 keV and 40-600 keV bands, respectively. XIS-0, XIS-2, and XIS-3 are front-side-illuminated CCDs, and XIS-1 is the back-side-illuminated one. In this paper, we analyze the data of the XIS-0, XIS-1, XIS-3 because XIS-2 became inoperable since 2007 November due to a detector trouble and HXD was not usable due to the power shortage of the spacecraft during our observation epochs.

\section{Analysis and Results}
We analyze the Suzaku data in a standard manner, using HEAsoft version 6.17 and calibration database released on 2015 March 12. For the XIS data, we analyze the version 2.3 cleaned events distributed by the Suzaku pipeline processing team. To extract the light curves and spectra, we set the source region as a circle around the detected position with a radius of 100 arcsec. The background for the XIS data is taken from a source-free region in the field of view with an approximately same offset angle from the optical axis as the source.

\subsection{Light Curves}
\begin{figure}
\begin{center}
\FigureFile(80mm,60mm){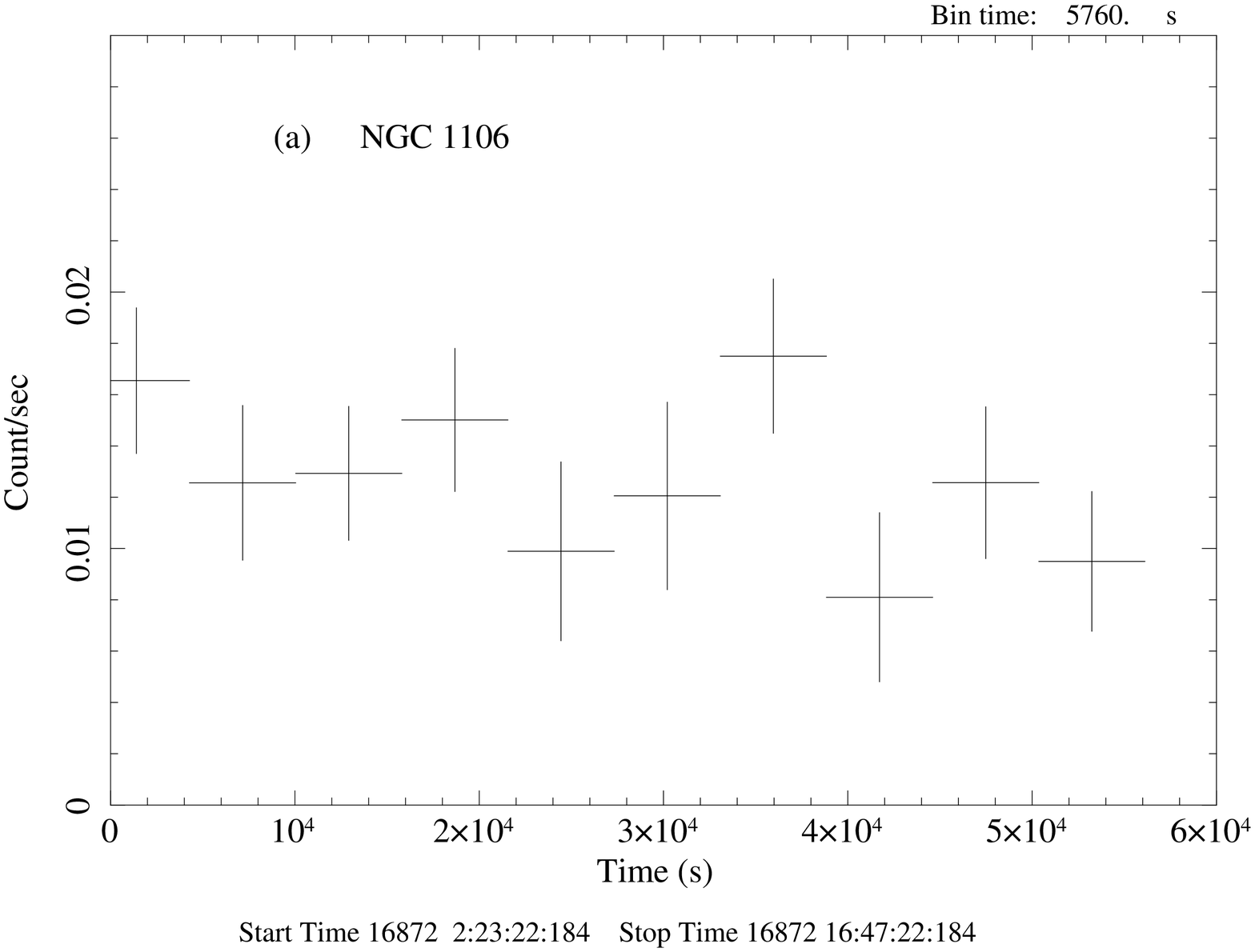}
\FigureFile(80mm,60mm){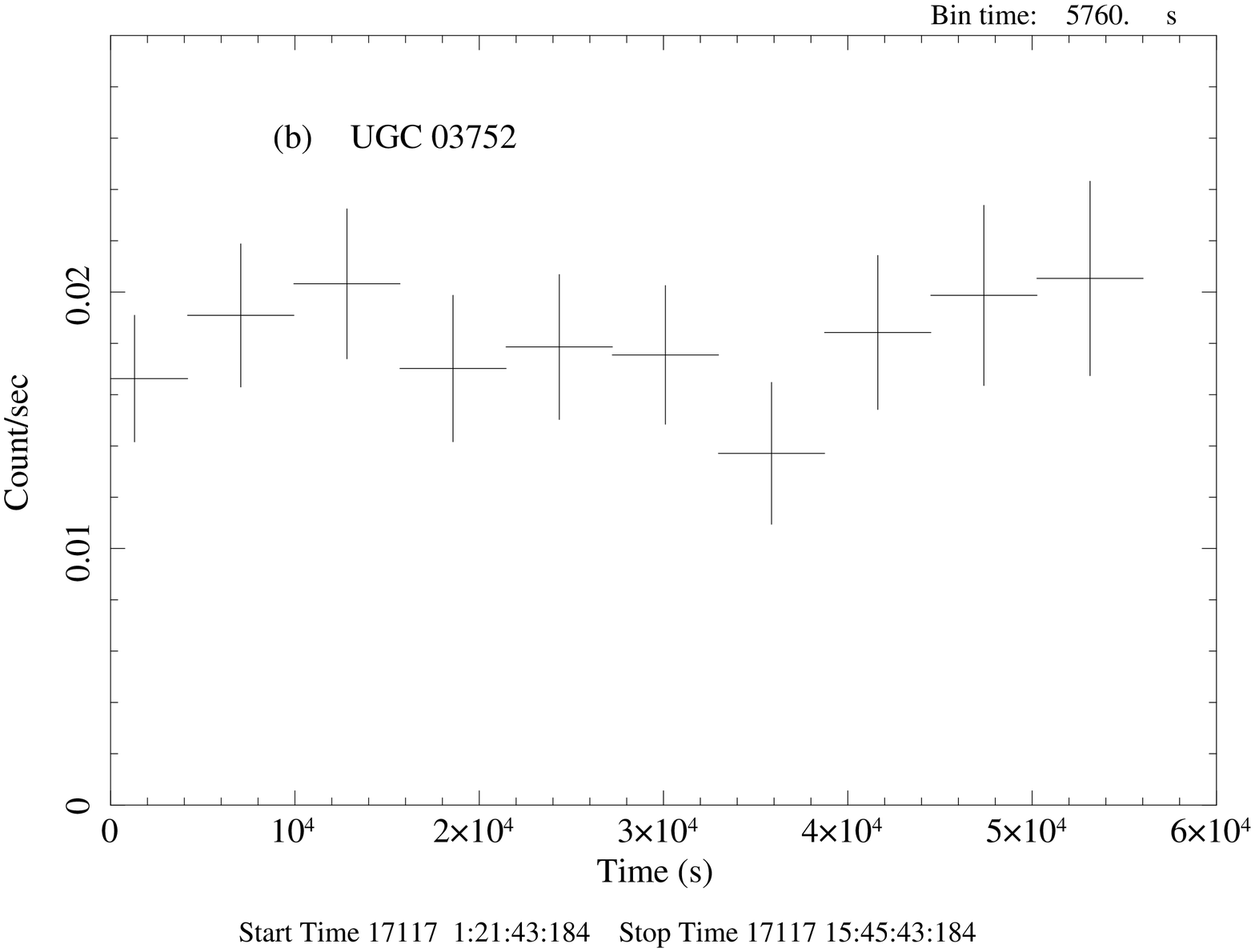}
\FigureFile(80mm,60mm){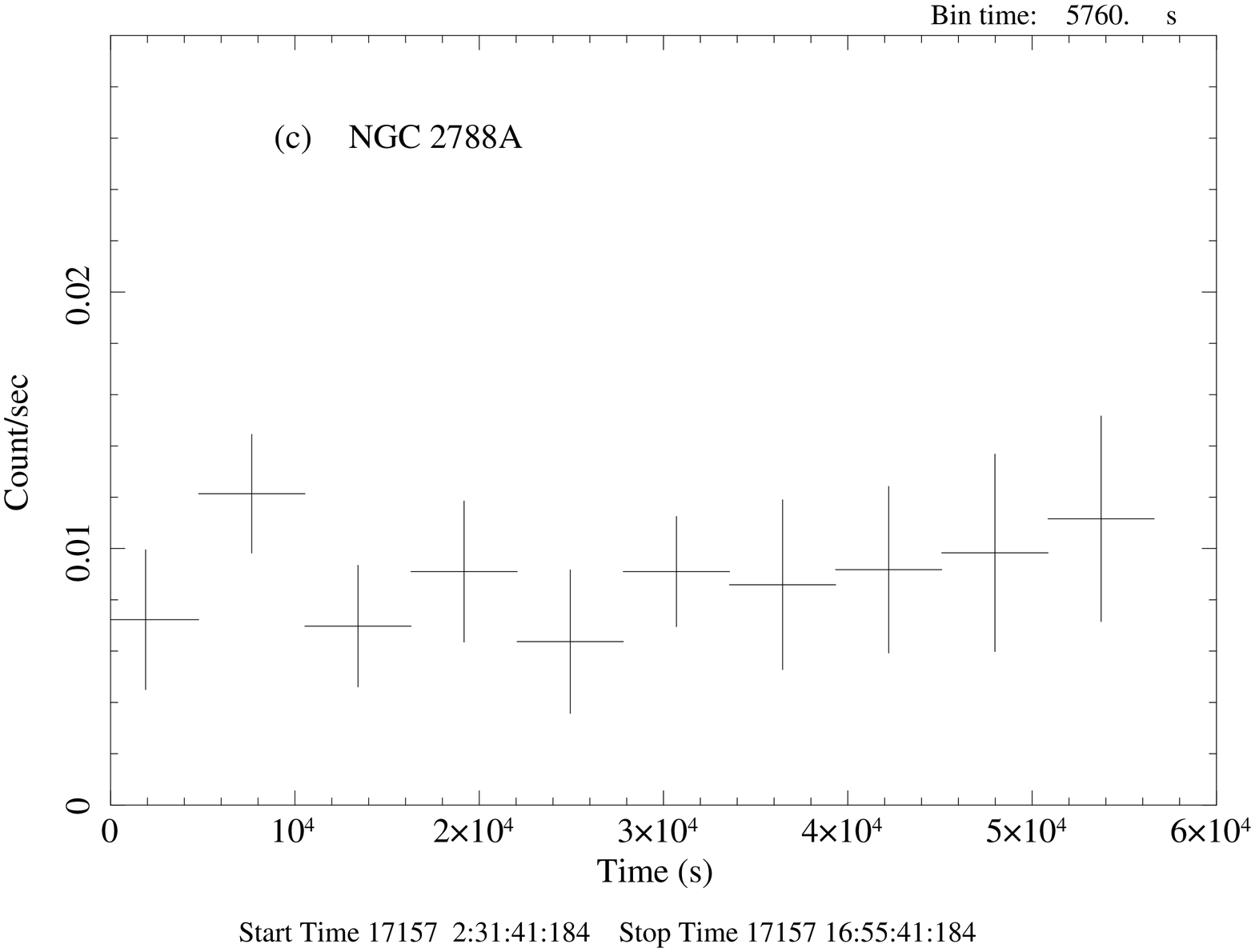}
\end{center}
\caption{
Background-subtracted light curves obtained with the XIS (average of XIS-0, XIS-1, and XIS-3) in the 2--10 keV band. The bin size is 96 minutes.}
\end{figure}

Figure~1 shows the background-subtracted light curves of our targets obtained with the XIS in the 2--10 keV band. To minimize any systematic uncertainties caused by the orbital change of the satellite, we merge data taken during one orbit (96 minutes ) into one bin. Then, to check if there is any significant time variability during the observations, we perform a simple $\chi^2$ test to each light curve assuming a null hypothesis of a constant flux. We detect no significant time variability on a timescale longer than 96 minutes for all the targets. Thus, we analyze the spectrum averaged over the whole observation.

\subsection{Spectra}
For spectral analysis, we use the data of the BI-XIS (XIS-1) and FI-XISs (XIS-0 + XIS-3) in the energy band of 0.5--8.0 keV and 2.0--10 keV, respectively; the data of the FI-XISs below 2 keV are discarded because all targets show little counts owning to the lower quantum efficiency than that of the BI-XIS. To obtain better constraint on the hard X-ray spectra, we also use the 70-month averaged Swift/BAT spectra in the 15--100 keV band. The 1.7--1.9 keV band of the BI-XIS spectrum is excluded to avoid systematic uncertainties in the energy response around the Si K-edge region. The spectra of the three targets folded with the energy responses are shown in Figure~2.  The presence of prominent iron-K emission line are noticeable.

We simultaneously fit the Suzaku and Swift/BAT spectra, which totally cover the 0.5--100 keV band. The Galactic absorption is always included in spectral models by \textbf{phabs} with the hydrogen column density $N_{\mathrm{H}}^{\mathrm{Gal}}$ estimated from the $\mathrm{H_I}$ map \citep{Kalberla05}. To take into account the cross-calibration uncertainties in the absolute effective area between the BI-XIS and FI-XISs, a constant factor is multiplied to the model, which is fixed at unity for FI-XISs and is left free for the BI-XIS. Usually, to take into account possible time variability among different epochs of observations, the cross normalization between Suzaku and BAT data must be introduced, at least for the transmitted component (see e.g., \cite{Tazaki13}). In our case, however, we find that Suzaku data are consistent with no time variability compared with the 70-month averaged Swift/BAT fluxes for all targets within the statistical error. In fact, as we show later, a large fraction of the X-ray flux below 10 keV in NGC 1106 and NGC 2788A is likely attributable to the reflection component from the torus, which is expected to show little variability on timescales shorter than years if the size of the emitting region is a parsec scale. Hence, we fix the constant factor for the Swift/BAT spectrum at unity in our spectral analysis. We confirm that the main results (such as line-of-sight column densities) are not significantly affected by possible time variability in the transmitted component.

\begin{figure}
\begin{center}
\FigureFile(80mm,60mm){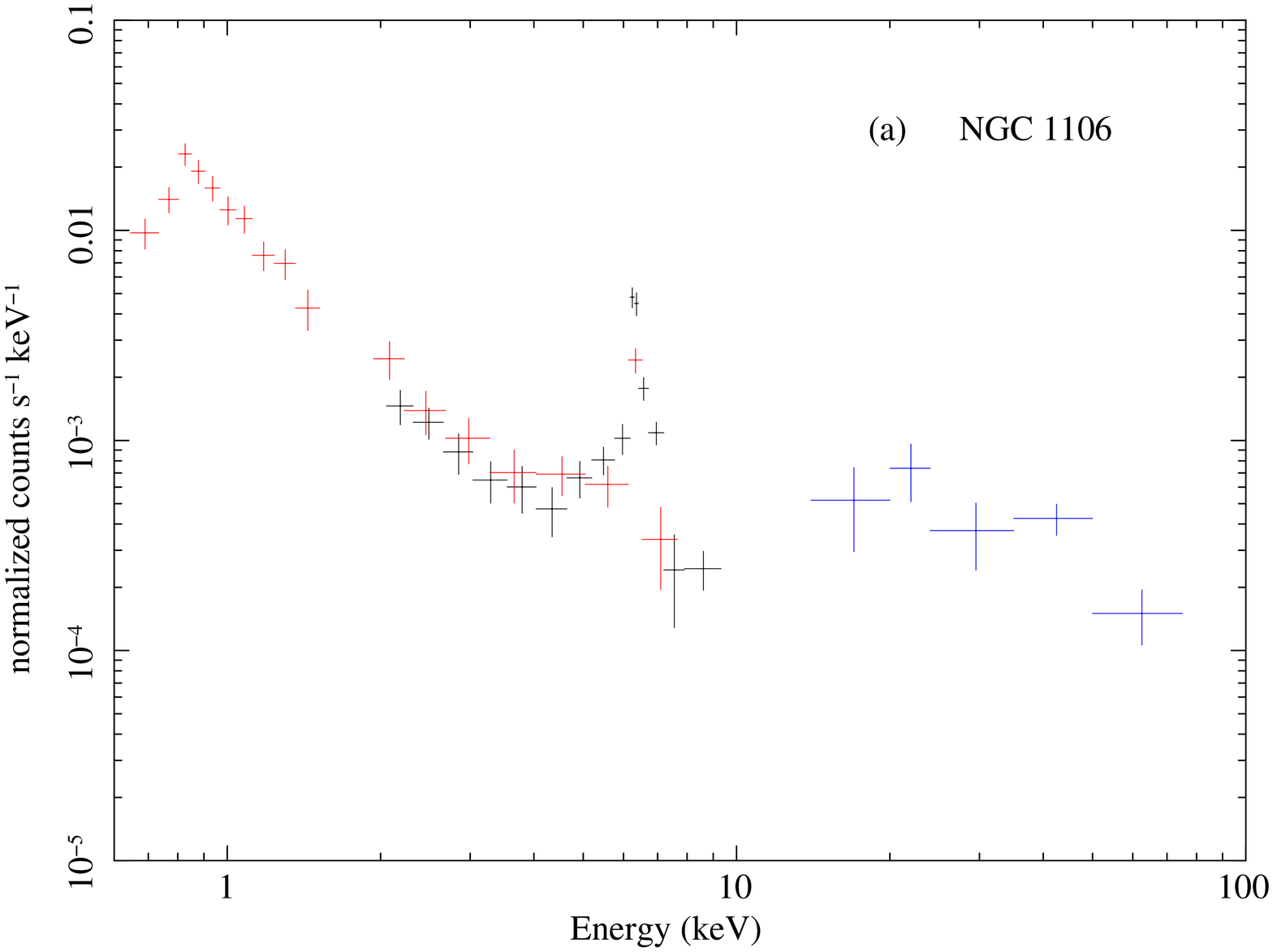}
\FigureFile(80mm,60mm){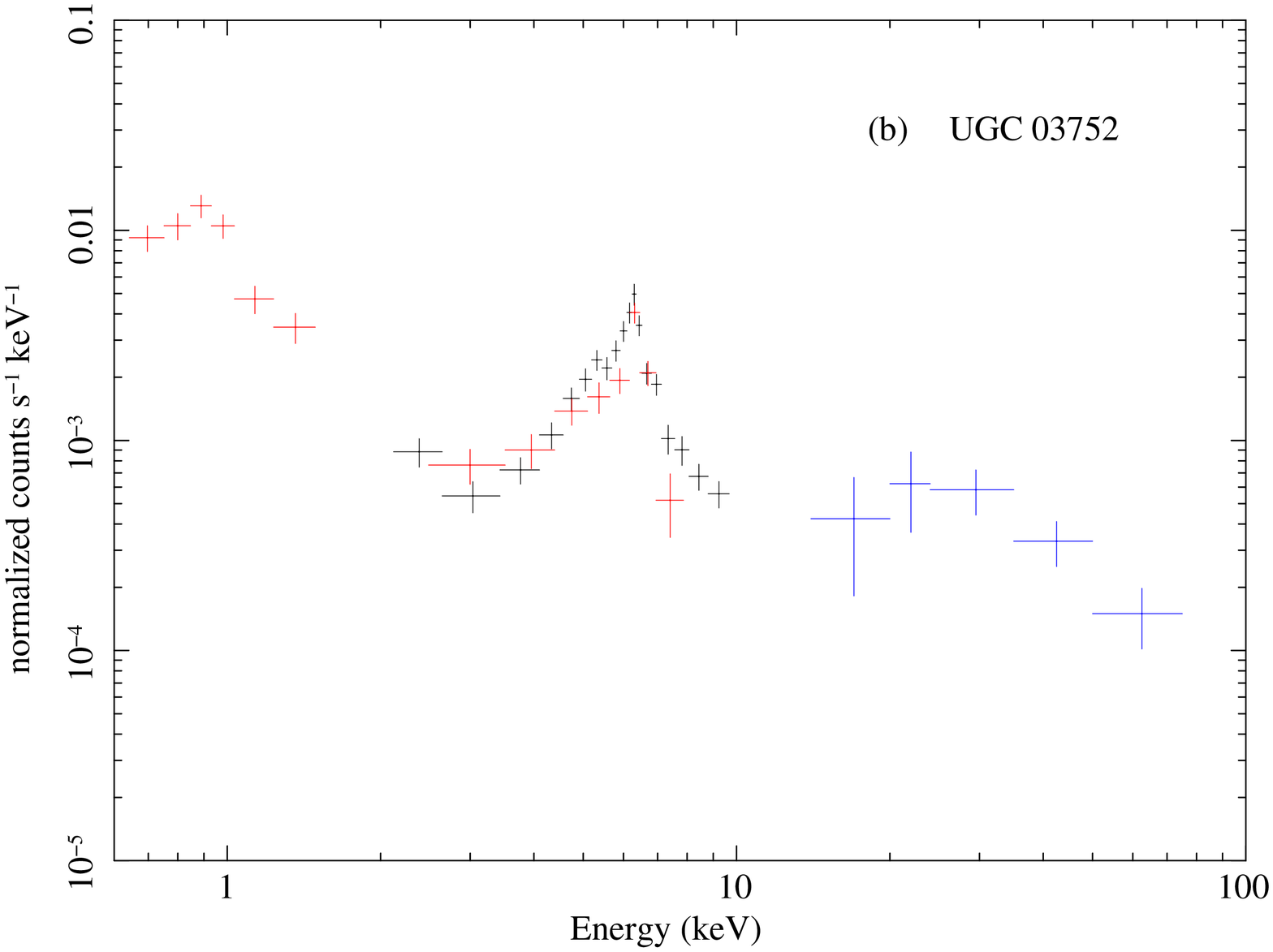}
\FigureFile(80mm,60mm){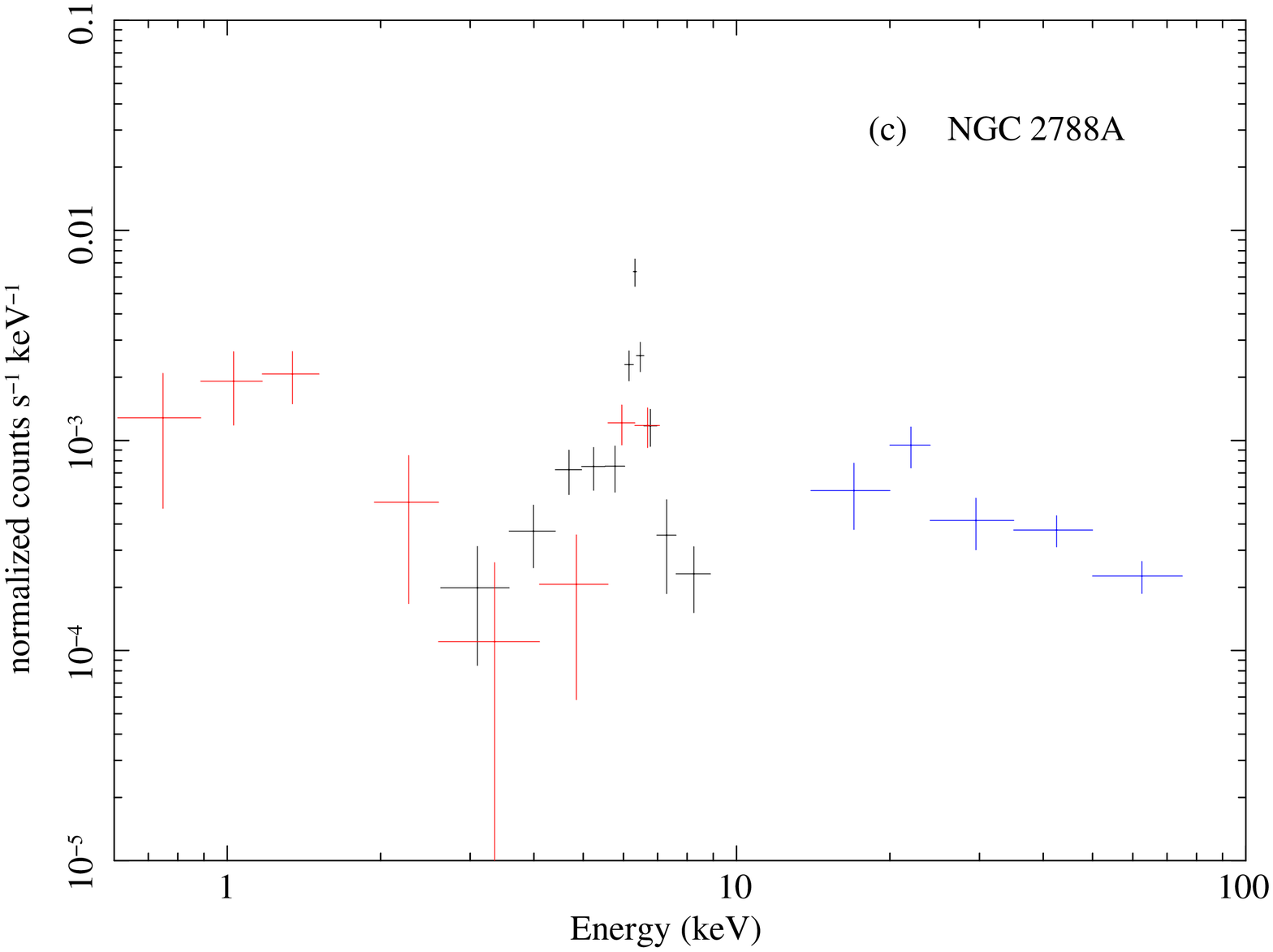}
\end{center}
\caption{
The observed XIS spectra folded with the energy response and the Swift/BAT spectra in units of photon cm$^{-2}$ ks$^{-1}$  keV$^{-1}$. (a) NGC 1106. (b) UGC 03752. (c) NGC 2788A. The black crosses, red crosses, and blue crosses represent the data of FI-XIS, BI-XIS, and BAT, respectively.}
\end{figure}

\subsubsection{Conventional Model}
\begin{table*}
\caption{Best-fit parameters with conventional model}
\begin{center}
\begin{tabular}{lcccc}
\hline
\multicolumn{1}{c}{} & Parameter & NGC 1106 & UGC 03752 & NGC 2788A			\\
\hline
(1)	& $X_{\mathrm{norm}}$
& $1.01_{-0.14}^{+0.15}$	& $1.06_{-0.10}^{+0.11}$	& $0.96_{-0.22}^{+0.25}$\\
(2)	& $N_{\mathrm{H}}$ 	[10$^{24}$ cm$^{-2}$]
& $1.36_{-0.20}^{+0.21}$	& $0.50_{-0.07}^{+0.07}$	& $0.99_{-0.26}^{+0.29}$\\
(3)	& $\Gamma$
& $1.79_{-0.22}^{+0.23}$	& $1.56_{-0.15}^{+0.17}$	& $1.40_{-0.30}^{+0.28}$\\
(4)	& $P_{\mathrm{norm}}$	[10$^{-2}$ keV$^{-1}$ cm$^{-2}$ s$^{-1}$]
& $0.13_{-0.08}^{+0.17}$	& $0.05_{-0.02}^{+0.03}$	& $0.03_{-0.02}^{+0.07}$\\
(5)	& $f_{\mathrm{scat}}$	[\%]
& $2.93_{-1.32}^{+2.42}$	& $3.85_{-1.03}^{+1.34}$	& $3.11_{-1.79}^{+4.40}$\\
(6)	& $E_{\mathrm{cent}}$	[keV]
& $6.40_{-0.03}^{+0.02}$	& $6.38_{-0.03}^{+0.04}$	& $6.40_{-0.02}^{+0.02}$\\
(7)	& $F_{\mathrm{norm}}$ 	[10$^{-5}$ cm$^{-2}$ s$^{-1}$]
& $0.58_{-0.10}^{+0.10}$	& $0.42_{-0.10}^{+0.10}$	& $0.65_{-0.15}^{+0.15}$\\
(8)	& $kT$ 					[keV]
& $0.71_{-0.09}^{+0.07}$	& $0.74_{-0.13}^{+0.08}$	& 					\\
(9)	& EW					[keV]
& 0.95					& 0.31					& 1.20				\\
& $\chi^2/$dof
& 23.6/29					& 28.3/29					& 18.1/17				\\
\hline
\multicolumn{1}{@{}l@{}}{\hbox to 0pt{\parbox{150mm}
{\footnotesize
\textbf{Notes.}	\\
(1)The normalization of BI-XIS relative to FI-XISs.\\
(2)The hydrogen column density.\\
(3)The photon index of the power law.	\\
(4)The normalization of the power law.\\
(5)The scattering fraction relative to the flux measured with Swift/BAT.\\
(6)The center energy of the iron-K emission line.\\
(7)The normalization of the iron-K emission line.\\
(8)The temperature of \textbf{apec} model.\\
(9)The equivalent width of the iron-K emission line with respect to the whole continuum.\\
}\hss}}
\end{tabular}
\end{center}
\end{table*}

For consistency with previous works, we first consider an analytic model often adopted to represent the broadband spectra of obscured AGNs (e.g., \cite{Eguchi09}). In XSPEC terminology, the model is expressed as
\begin{eqnarray}
\mathrm{model1}
& = & \mathbf{zphabs*zpowerlw*zhighect}	\nonumber\\
& + & \mathbf{const*zpowerlw*zhighect}		\nonumber\\
& + & \mathbf{zphabs*pexrav+zgauss+apec}.
\end{eqnarray}
This model is composed of five components: (1) The first term is the transmitted component absorbed by cold matter at the source redshift (\textbf{zphabs}). It is known that the intrinsic spectra of AGNs are well approximated by a power law (\textbf{zpowerlw}) with an exponential cutoff (\textbf{zhighect}). Since we cannot constrain the cutoff energy for all targets, we fix it at 360 keV for consistency with the torus  model by \citet{Ikeda09}. (2) The second term is the scattered component. We multiply a scattering fraction $f_{\mathrm{scat}}$ (\textbf{const}) to the same cutoff power law without absorption as the first term. (3) The third term is the reflection component. We assume the same absorption (\textbf{zphabs}) as that for the transmitted component. We adopt the analytic model by \citet{Magdziarz95} (\textbf{pexrav}), assuming Compton reflection from optically thick, cold matter. The relative reflection intensity is defined as $R \equiv \Omega/2\pi$, where $\Omega$ is the solid angle of the reflector. The inclination angle is fixed at 60 degrees. The photon index and normalization of the cutoff power law are linked to those of the transmitted component. (4) The fourth term is the iron-K emission line around 6.4 keV (\textbf{zgauss}). We fix the line width at 10 eV but set the central energy free. (5) The fifth term represents emission from an optically thin thermal plasma (\textbf{apec}) in the host galaxy \citep{Smith01}, which is often observed below 1 keV (e.g.,\cite{Terashima02}).

We find this model can well reproduce the observed broadband spectra in terms of $\chi^2$ values. The best-fit parameters are summarized in Table~2. Since the reflection strength cannot be well constrained from our data, we fix $R=1$. In the case of NGC 2788A, the last term (the soft component) is not required. We obtain the line-of-sight column densities of $N_{\mathrm{H}} \gtrsim 10^{24}$ cm$^{-2}$ in NGC 1106 and NGC 2788A. This is consistent with their large equivalent width of the narrow iron-K emission line of $\sim$1 keV.

\subsubsection{Torus Models}
While the above conventional model utilizing the \textbf{pexrav} code for cold reflection has been often used to fit the X-ray spectra of obscured AGNs, the actual geometry of the reflector is expected to be quite different from that assumed in the \textbf{pexrav} model (i.e., a semi-infinite slab of optically thick material). Also, there is no guarantee that the reflector in AGNs can be regarded to be optically thick for Compton scattering. In this sense, this model may not be physically self-consistent.

Monte Carlo simulations are very useful approach to investigate the spectra of heavily obscured AGNs with complex geometry of surrounding matter. Here we apply two major Monte-Carlo based numerical spectral models of AGN torus with different geometry, so-called the MYTorus model \citep{Murphy09} and the Ikeda torus model \citep{Ikeda09}, to our broadband X-ray spectra. Both models assume cold matter with a uniform density for the torus (``smooth torus'') and isotropic irradiation from the central source with a power-law spectrum (with an exponential cutoff in the Ikeda torus model).

The MYTorus model considers a tube-like symmetric matter (see figure 1 in \cite{Murphy09}), corresponding to a classical doughnut-type geometry for the obscuring torus. The half opening angle of the torus is fixed at 60 degrees. Besides the normalization and photon index of the incident spectrum, it has two parameters, the hydrogen column density along the equatorial plane $N_{\mathrm{H}}^{\mathrm{Eq}}$ and the inclination angle $\theta_{\mathrm{incl}}$ between the line-of-sight and the symmetry axis of the torus (i.e., $\theta_{\mathrm{incl}} = 90$ corresponds to the edge-on case). The line-of-sight column density $N_{\mathrm{H}}^{\mathrm{LS}}$ in the case of $\theta _{\mathrm{incl}}
\geq 60$ is related to $N_{\mathrm{H}}^{\mathrm{Eq}}$ and $\theta_{\mathrm{incl}}$ as
\begin{equation} 
N_{\mathrm{H}}^{\mathrm{LS}} = 
(1-4\cos^2\theta_{\mathrm{incl}}) N_{\mathrm{H}}^{\mathrm{Eq}} \quad (\theta_{\mathrm{incl}} \geq 60).
\end{equation}
From all targets, we obtained acceptable fits with the spectral model utilizing MYTorus (hereafter model~2), which is expressed in XSPEC terminology as:
\begin{eqnarray}
\mathrm{model2}
& = & \mathbf{etable[mytorus\_Ezero\_v00.fits]*zpowerlw}	\nonumber\\
& + & \mathbf{const*zpowerlw*zhighect}					\nonumber\\
& + & \mathbf{atable[mytorus\_scatteredH500\_v00.fits]}	\nonumber\\
& + & \mathbf{atable[mytl\_V000010nEp000H500\_v00.fits]}	\nonumber\\
& + & \mathbf{apec}.
\end{eqnarray}
The model is composed of five components: (1) the absorbed transmitted component, (2) the scattered component, (3) the reflection component from the torus, (4) the iron-K emission line from the torus, and (5) the optically-thin thermal component from the host galaxy (only for NGC 1106 and UGC 03752). The best-fit parameters are summarized in Table~3. Figure~4 (a)-(c) plots the best-fit model in units of $E F_{\mathrm{E}}$, where $F_{\mathrm{E}}$ is the energy flux at the energy $E$, with separate model components.

The Ikeda torus model assumes a nearly spherical geometry with two holes of conical shape along the polar axis (see figure 2 in \cite{Ikeda09}), similar to that adopted by \citet{Brightman11a, Brightman11b}. This model has three parameters on the torus structure, the hydrogen column density along the equatorial plane $N_{\mathrm{H}}^{\mathrm{Eq}}$, the inclination angle of the torus $\theta_{\mathrm{incl}}$, and the half opening angle of the torus $\theta_{\mathrm{open}}$. The ratio between the inner radius $r_{\mathrm{in}}$ and outer radius $r_{\mathrm{out}}$ is fixed at 0.01. A power law with a cutoff energy of 360 keV is assumed as the spectrum of the primary emission. Although the half opening angle can be set as a free parameter unlike in the MYTorus model, we find that it cannot be constrained from our data owning to the limited photon statistics. Hence, we fix $\theta_{\mathrm{open}}$ at 60 degrees, which enables us to directly compare with the MYTorus results. The hydrogen column density along the line-of-sight $N_{\mathrm{H}}^{\mathrm{LS}}$ in the case of $\theta_{\mathrm{incl}} \geq \theta_{\mathrm{open}}$ is determined as
\begin{equation}
N_{\mathrm{H}}^{\mathrm{LS}}
= \frac{0.01(\cos\theta_{\mathrm{incl}}-\cos\theta_{\mathrm{open}}) +\sin(\theta_\mathrm{incl}-\theta_{\mathrm{open}})}
{0.99(0.01\cos\theta_{\mathrm{incl}}
+\sin(\theta_{\mathrm{incl}}-\theta_{\mathrm{open}}))}N_{\mathrm{H}}^{\mathrm{Eq}}.
\end{equation}
We also obtain acceptable fits for all the spectra with the following model utilizing the Ikeda torus model (model~3).
\begin{eqnarray}
\mathrm{model3}
& = & \mathbf{torusabs*zpowerlw*zhighect}		\nonumber\\
& + & \mathbf{const*zpowerlw*zhighect}			\nonumber\\
& + & \mathbf{atable[refl1\_torus.fits]}			\nonumber\\
& + & \mathbf{atable[refl2\_torus.fits]}			\nonumber\\
& + & \mathbf{atable[refl\_fe\_torus.fits]+apec}.
\end{eqnarray}
Model 3 has six components: (1) the absorbed transmitted component, (2) the scattered component, (3) the unabsorbed reflection component from the far-side torus, (4) the reflection components absorbed by the near-side torus, (5) the iron-K emission line from the torus, and (6) the optically-thin thermal component from the host galaxy (only for NGC 1106 and UGC 03752). Table~4 summarizes the best-fit parameters. The best-fit models are plotted in Figure~4 (d)-(f) for the 3 targets, and the unfolded spectra in units of $EF_{\mathrm{E}}$ are shown in Figure~3.

\begin{figure}
\begin{center}
\FigureFile(80mm,60mm){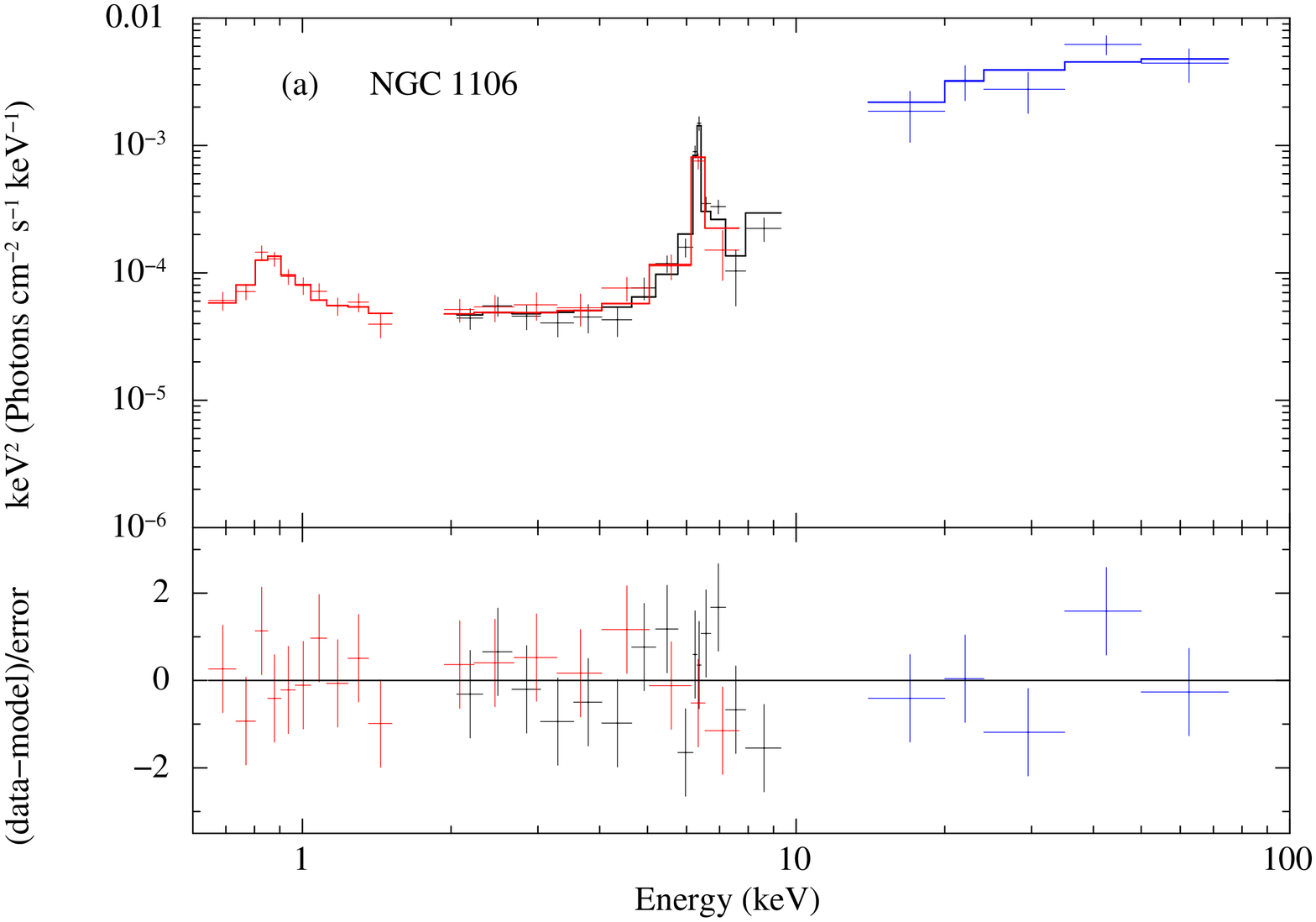}
\FigureFile(80mm,60mm){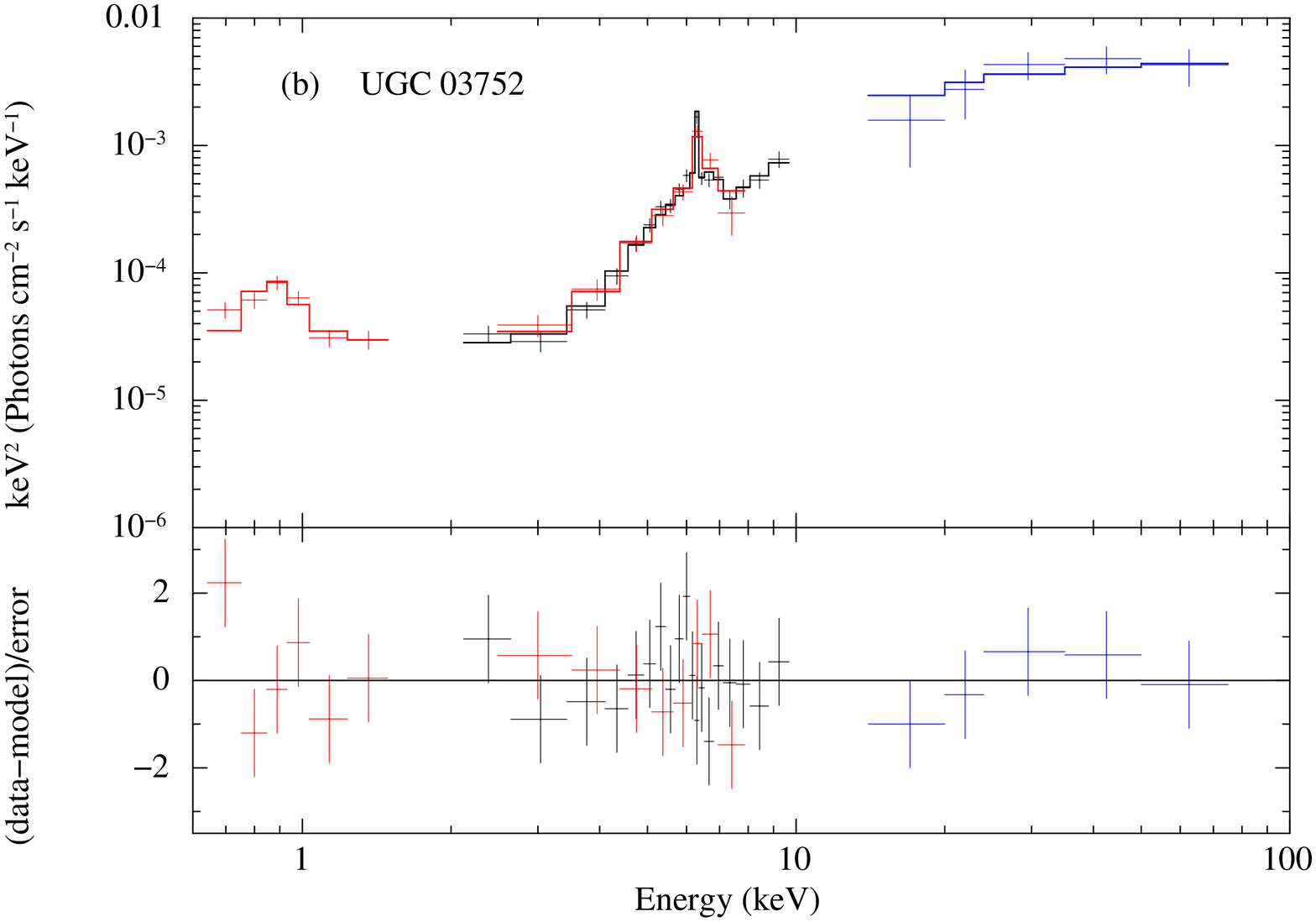}
\FigureFile(80mm,60mm){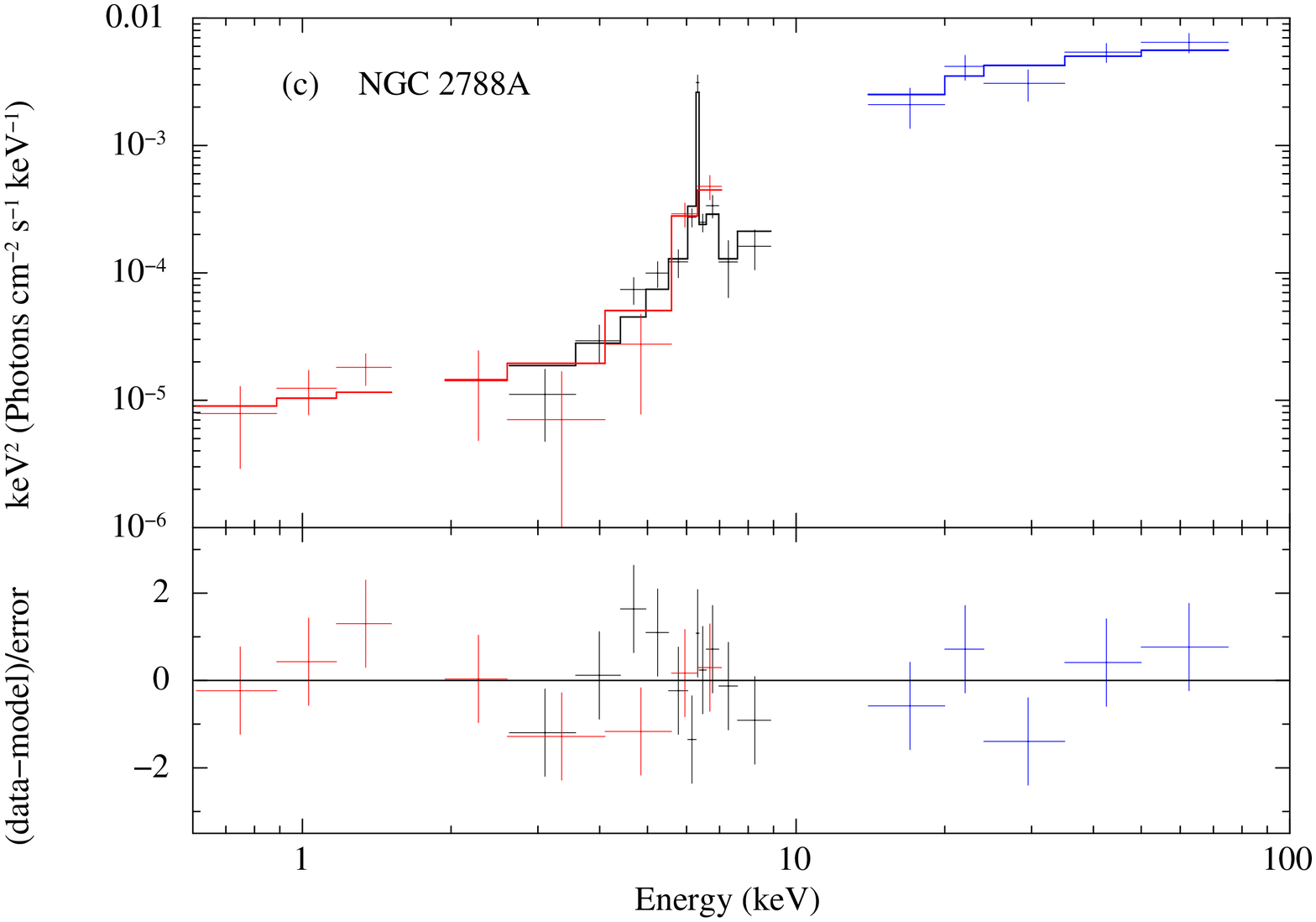}
\end{center}
\caption{
The unfolded spectra in units of $EF_{\mathrm{E}}$ fitted with model~3 (Ikeda torus model). (a) NGC 1106. (b) UGC 03752. (c) NGC 2788A. The black crosses, red crosses, and blue crosses represent the data of  FI-XISs, BI-XIS, and BAT, respectively. The best-fit models are plotted with solid lines.}
\end{figure}

\begin{table*}
\caption{Best-fit parameters with MYTorus model}
\begin{center}
\begin{tabular}{lcccc}
\hline
\multicolumn{1}{c}{} & Parameter & NGC 1106 & UGC 03752 & NGC 2788A\\
\hline
(1)	& $X_{\mathrm{norm}}$												
& $1.01_{-0.14}^{+0.15}$	& $1.07_{-0.10}^{+0.11}$	& $0.95_{-0.22}^{+0.25}$\\
(2)	& $N_{\mathrm{H}}^{\mathrm{Eq}}$	[10$^{24}$ cm$^{-2}$]		
& $2.33_{-0.40}^{+0.43}$	& $4.97_{-0.46}^{+0.47}$	& $2.11_{-0.55}^{+0.59}$\\
(3)	& $N_{\mathrm{H}}^{\mathrm{LS}}$	[10$^{24}$ cm$^{-2}$]		
& 2.13					& 0.18					& 1.71				\\
(4)	& $\theta_{\mathrm{incl}}$ 			[degree]							
& $81.6_{-4.1}^{+3.0}$		& $60.6_{-0.3}^{+9.7}$		& $77.5_{-10.8}^{+4.9}$	\\	
(5)	& $\Gamma$														
& $2.24_{-0.28}^{+0.26}$	& $1.75_{-0.09}^{+0.13}$	& $1.94_{-0.33}^{+0.36}$\\
(6)	& $P_{\mathrm{norm}}$				[10$^{-2}$ keV$^{-1}$ cm$^{-2}$ s$^{-1}$]
& $2.65_{-1.81}^{+5.66}$	& $0.15_{-0.06}^{+0.11}$	& $0.81_{-0.62}^{+2.99}$\\
(7)	& $f_{\mathrm{scat}}$				[\%]								
& $0.17_{-0.10}^{+0.26}$	& $1.24_{-0.26}^{+0.28}$	& $0.12_{-0.10}^{+0.36}$\\
(8)	& $kT$							[keV]							
& $0.75_{-0.08}^{+0.08}$	& $0.75_{-0.12}^{+0.08}$	&					\\
(9)	& EW							[keV]							
& 1.12					& 0.27					& 1.21				\\
	& $\chi^2/$dof														
& 28.3/30					& 32.2/30					& 23.0/18				\\
\hline
\multicolumn{1}{@{}l@{}}{\hbox to 0pt{\parbox{150mm}
{\footnotesize
\textbf{Notes.}\\
(1)The normalization of BI-XIS relative to FI-XISs.\\
(2)The hydrogen column density along the equatorial plane.\\
(3)The hydrogen column density along the line of sight.\\
(4)The inclination angle of the torus.\\
(5)The photon index of the power law.\\
(6)The normalization of the power law.\\
(7)The scattering fraction relative to the flux measured with Swift/BAT.\\
(8)The temperature of the \textbf{apec} model.\\
(9)The equivalent width of the iron-K emission line with respect to the whole continuum.\\
}\hss}}
\end{tabular}
\end{center}
\end{table*}

\begin{table*}
\caption{Best-fit parameters with Ikeda torus model}
\begin{center}
\begin{tabular}{lcccc}
\hline
\multicolumn{1}{c}{} & Parameter & NGC 1106 & UGC 03752 & NGC 2788A\\
\hline
(1)	& $X_{\mathrm{norm}}$												
& $1.00_{-0.14}^{+0.16}$	& $1.06_{-0.10}^{+0.11}$	& $0.95_{-0.22}^{+0.25}$\\
(2)	& $N_{\mathrm{H}}^{\mathrm{Eq}}$	[10$^{24}$ cm$^{-2}$]		
& $1.78_{-0.29}^{+0.25}$	& $3.55_{-0.34}^{+0.35}$	& $1.66_{-0.34}^{+0.68}$\\
(3)	& $N_{\mathrm{H}}^{\mathrm{LS}}$	[10$^{24}$ cm$^{-2}$]		
& 1.75					& 0.53					& 1.58				\\
(4)	& $\theta_{\mathrm{incl}}$			[degree]							
& $70.3_{-4.3}^{+9.4}$		& $60.1_{-0.1}^{+0.1}$		& $65.0_{-2.9}^{+6.8}$	\\	
(5)	& $\Gamma$														
& $1.92_{-0.22}^{+0.26}$	& $1.62_{-0.12}^{+0.18}$	& $1.64_{-0.14}^{+0.24}$\\
(6)	& $P_{\mathrm{norm}}$				[10$^{-2}$ keV$^{-1}$ cm$^{-2}$ s$^{-1}$]	& $0.63_{-0.39}^{+0.63}$	& $0.09_{-0.07}^{+0.03}$	& $0.22_{-0.10}^{+0.39}$\\
(7)	& $f_{\mathrm{scat}}$				[\%]								
& $0.67_{-0.24}^{+0.73}$	& $1.82_{-0.39}^{+0.41}$	& $0.53_{-0.31}^{+0.41}$\\
(8)	& $kT$							[keV]							
& $0.74_{-0.08}^{+0.07}$	& $0.75_{-0.12}^{+0.08}$	&					\\
(9)	& EW							[keV]							
& 0.98					& 0.33					& 1.25				\\
(10)	& $ F_{0.5-2}						[\mathrm{erg \ cm^{-2} \ s^{-1}}]$	
& $1.35 \times 10^{-13}$		& $8.07 \times 10^{-14}$		& $2.29 \times 10^{-14}$	\\
(11)	& $ F_{2-10}						[\mathrm{erg \ cm^{-2} \ s^{-1}}]$	
& $3.91 \times 10^{-13}$		& $7.08 \times 10^{-13}$		& $3.39 \times 10^{-13}$	\\
(12)	& $ F_{15-50} 						[\mathrm{erg \ cm^{-2} \ s^{-1}}]$	
& $7.20 \times 10^{-12}$		& $6.10 \times 10^{-12}$		& $7.60 \times 10^{-12}$	\\
(13)	& $ L_{2-10}						[\mathrm{erg \ s^{-1}}]$			
& $8.30 \times 10^{42}$		& $ 2.33 \times 10^{42}$		& $3.78 \times 10^{42}$	\\
	& $\chi^2/$dof														
& 28.0/30					& 28.2/30					& 18.6/18				\\
\hline
\multicolumn{1}{@{}l@{}}{\hbox to 0pt{\parbox{150mm}
{\footnotesize
\textbf{Notes.}\\
(1)The normalization of BI-XIS relative to FI-XISs\\
(2)The hydrogen column density along the equatorial plane.\\
(3)The hydrogen column density along the line of sight.\\
(4)The inclination angle of the torus.\\
(5)The photon index of the power law.\\
(6)The normalization of the power law.\\
(7)The scattering fraction relative to the flux measured with Swift/BAT.\\
(8)The temperature of the \textbf{apec} model.\\
(9)The equivalent width of the iron-K emission line with respect to the whole continuum.\\
(10)The observed flux with BI-XIS of Suzaku in the 0.5--2 keV band.\\
(11)The observed flux with FI-XISs of Suzaku in the 2--10 keV band.\\
(12)The observed flux with Swift/BAT in the 15--50 keV band.\\
(13)The intrinsic (de-absorbed) luminosity in the 2--10 keV band band.\\
}\hss}}
\end{tabular}
\end{center}
\end{table*}

\begin{figure*}
\begin{center}
\FigureFile(80mm,60mm){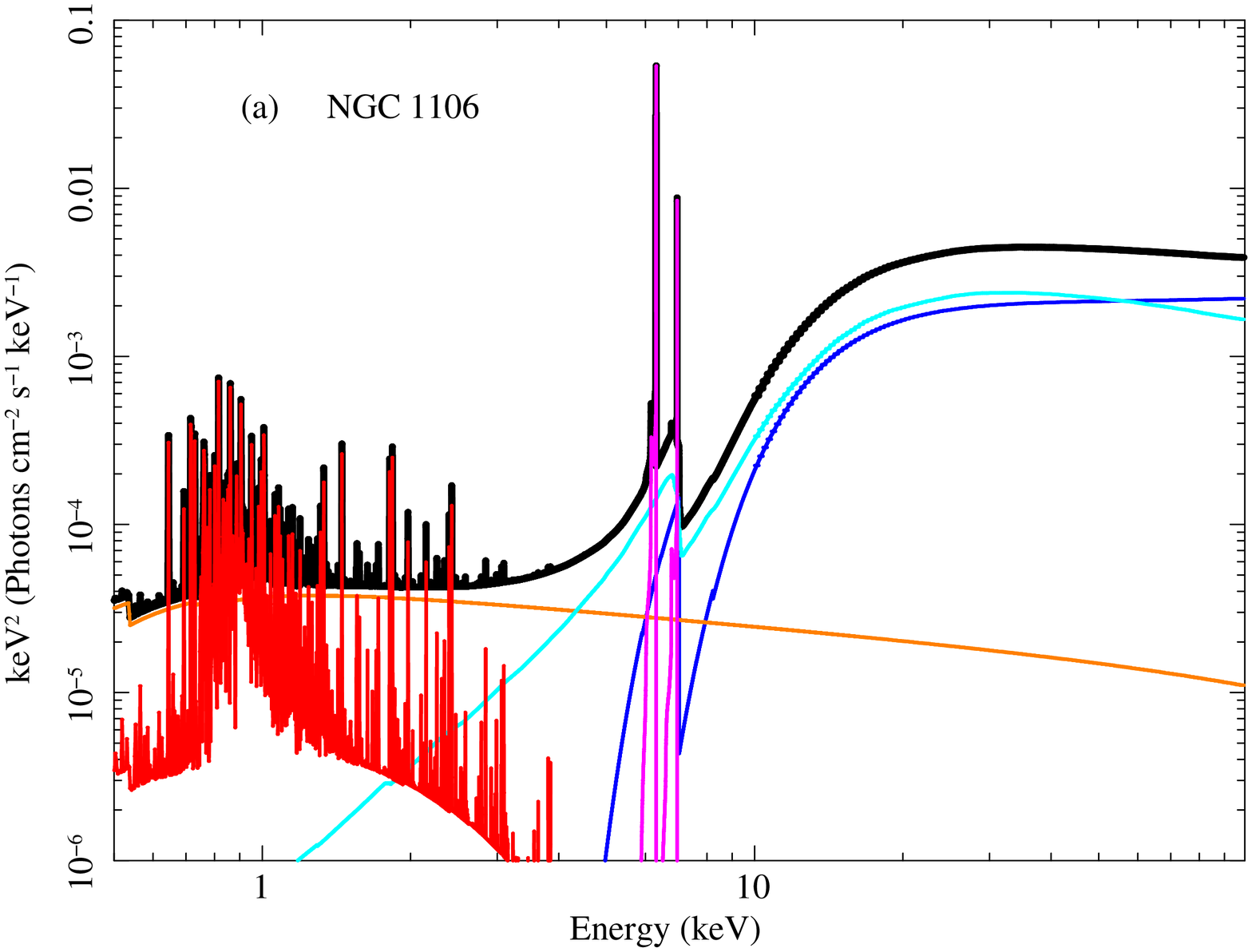}
\FigureFile(80mm,60mm){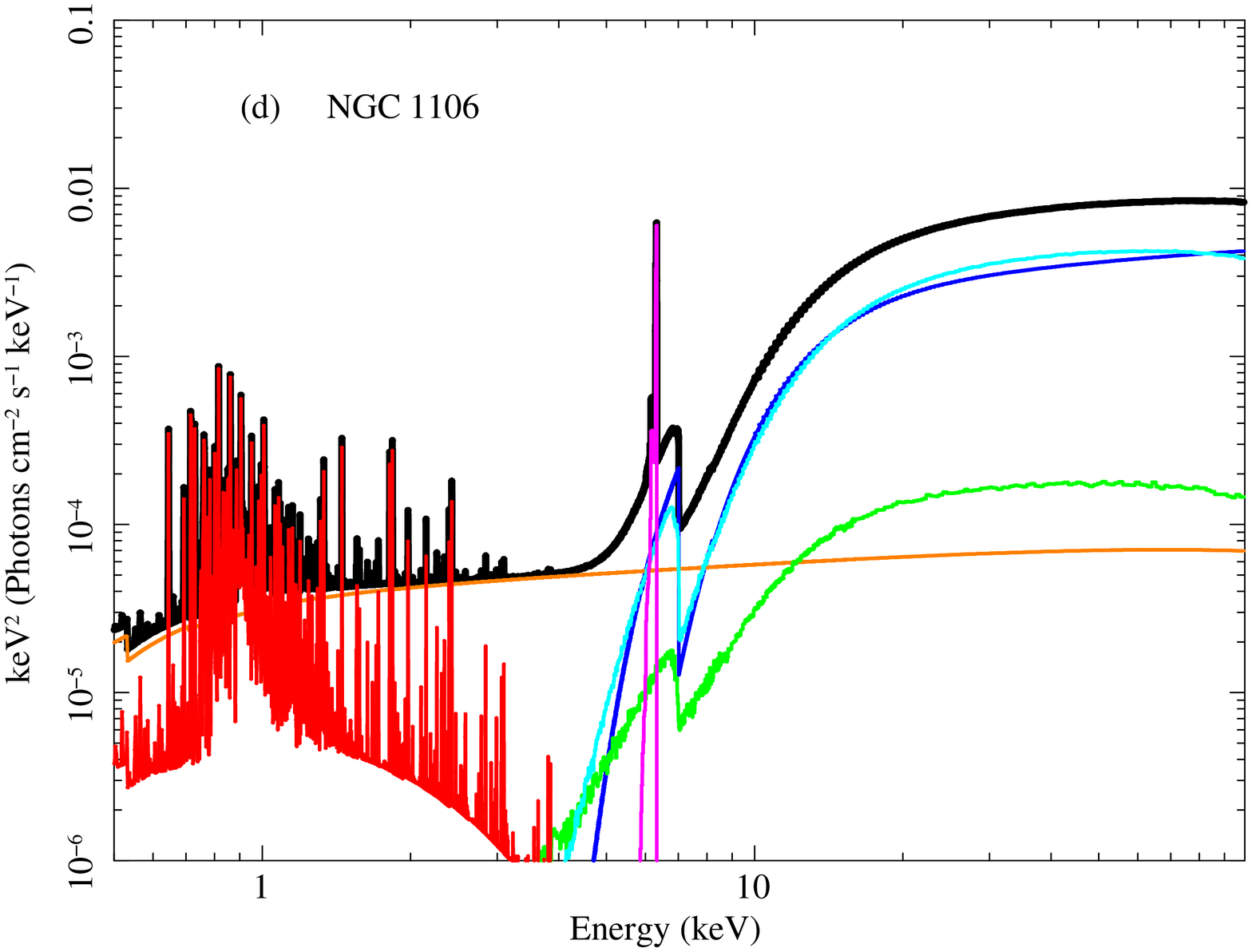}
\FigureFile(80mm,60mm){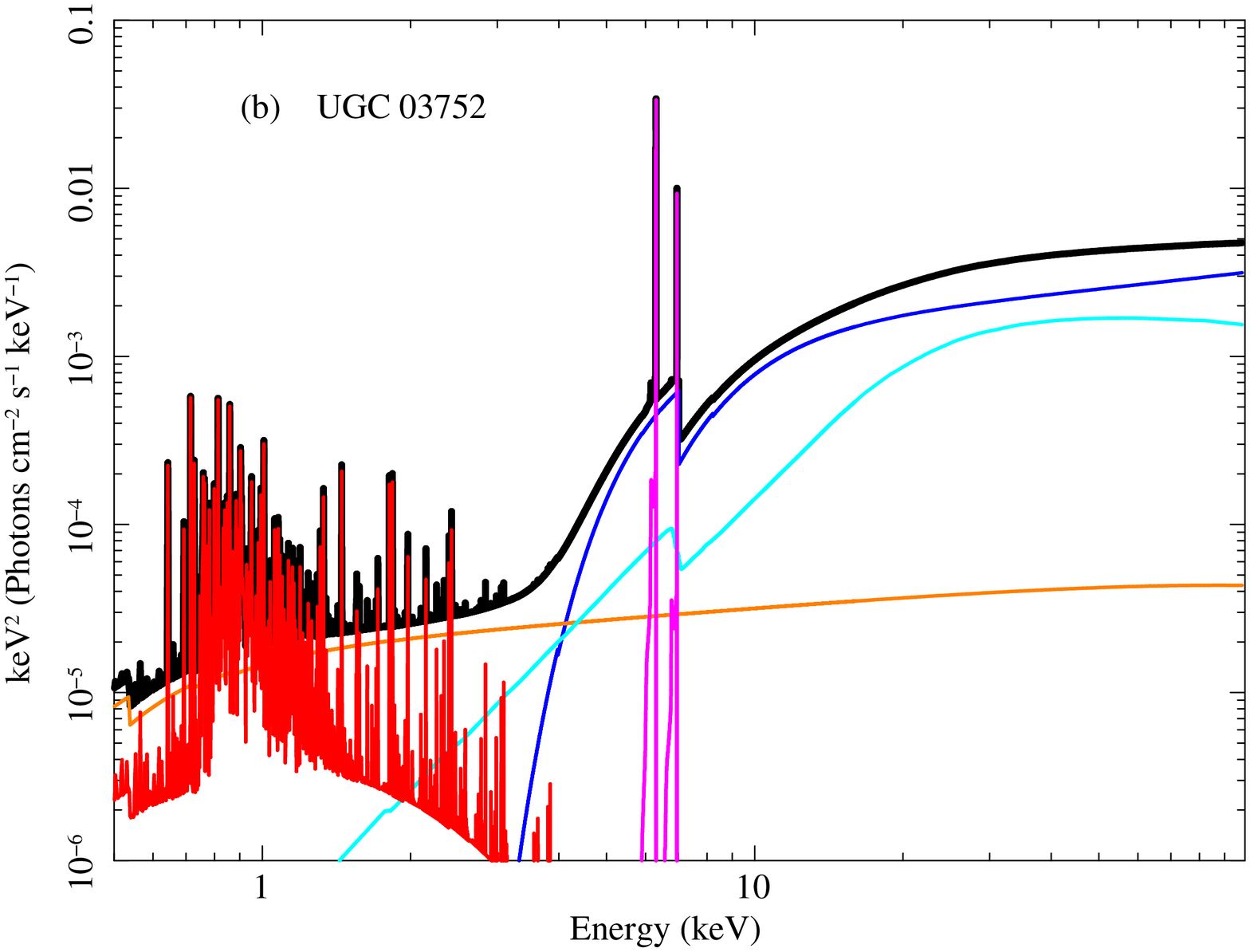}
\FigureFile(80mm,60mm){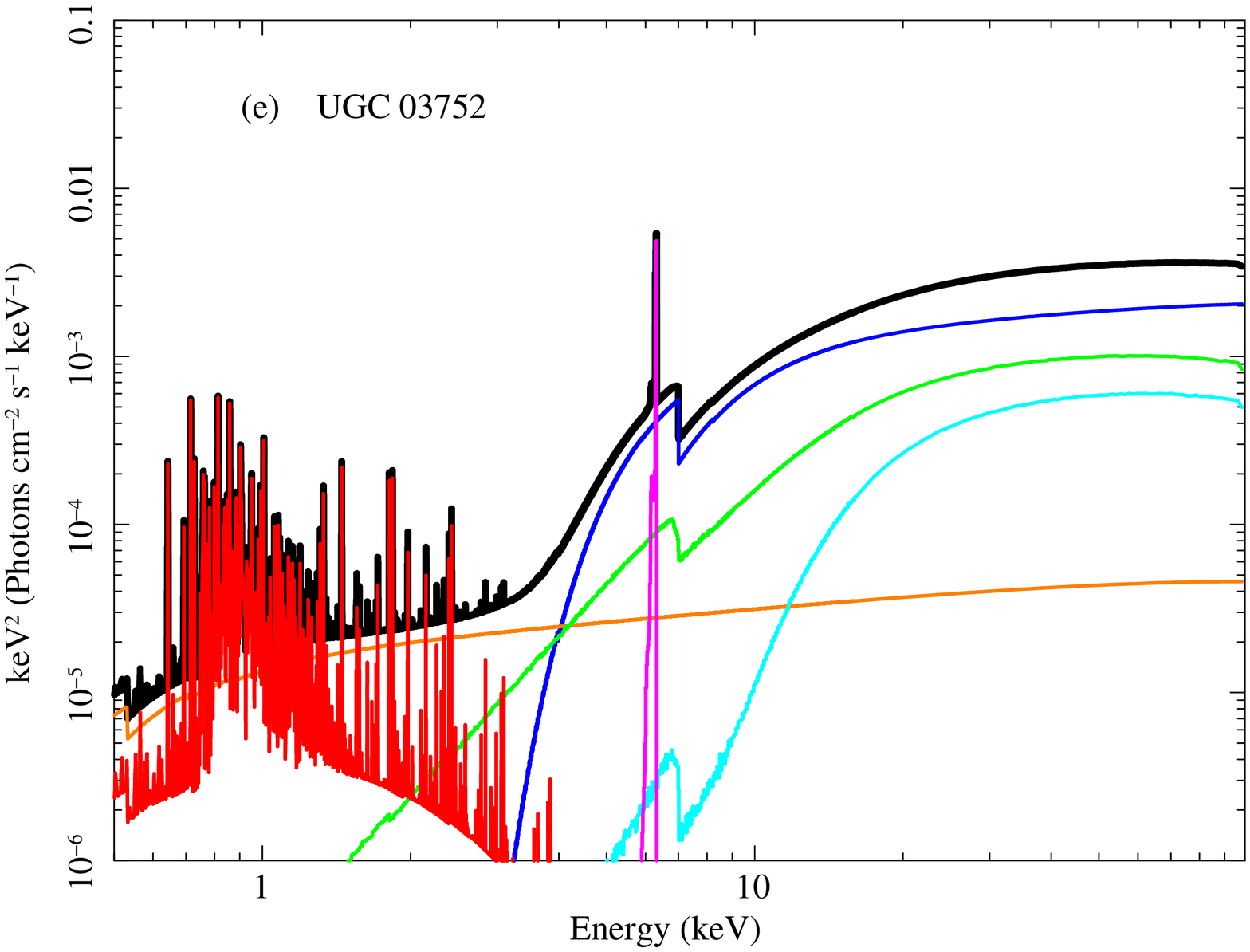}
\FigureFile(80mm,60mm){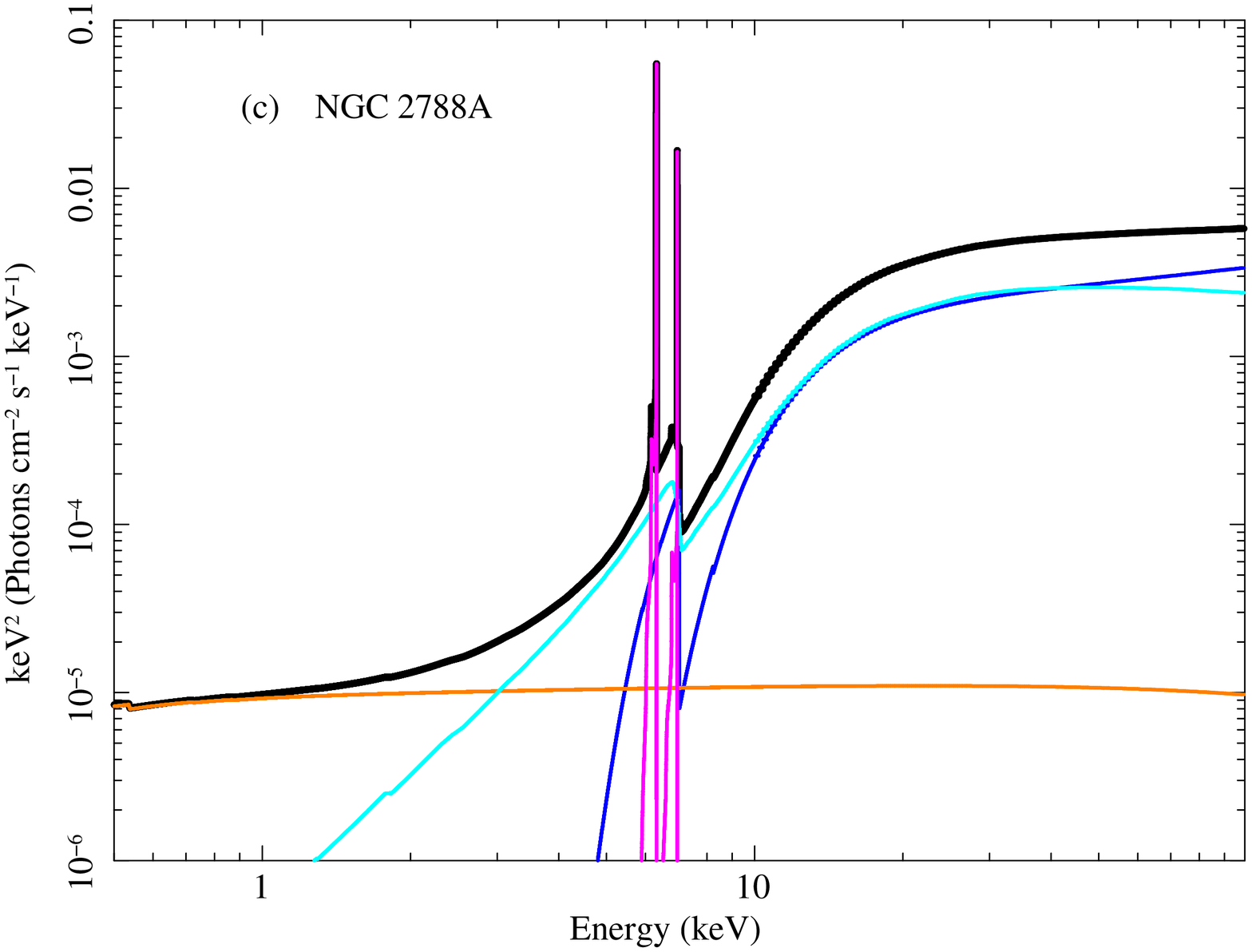}
\FigureFile(80mm,60mm){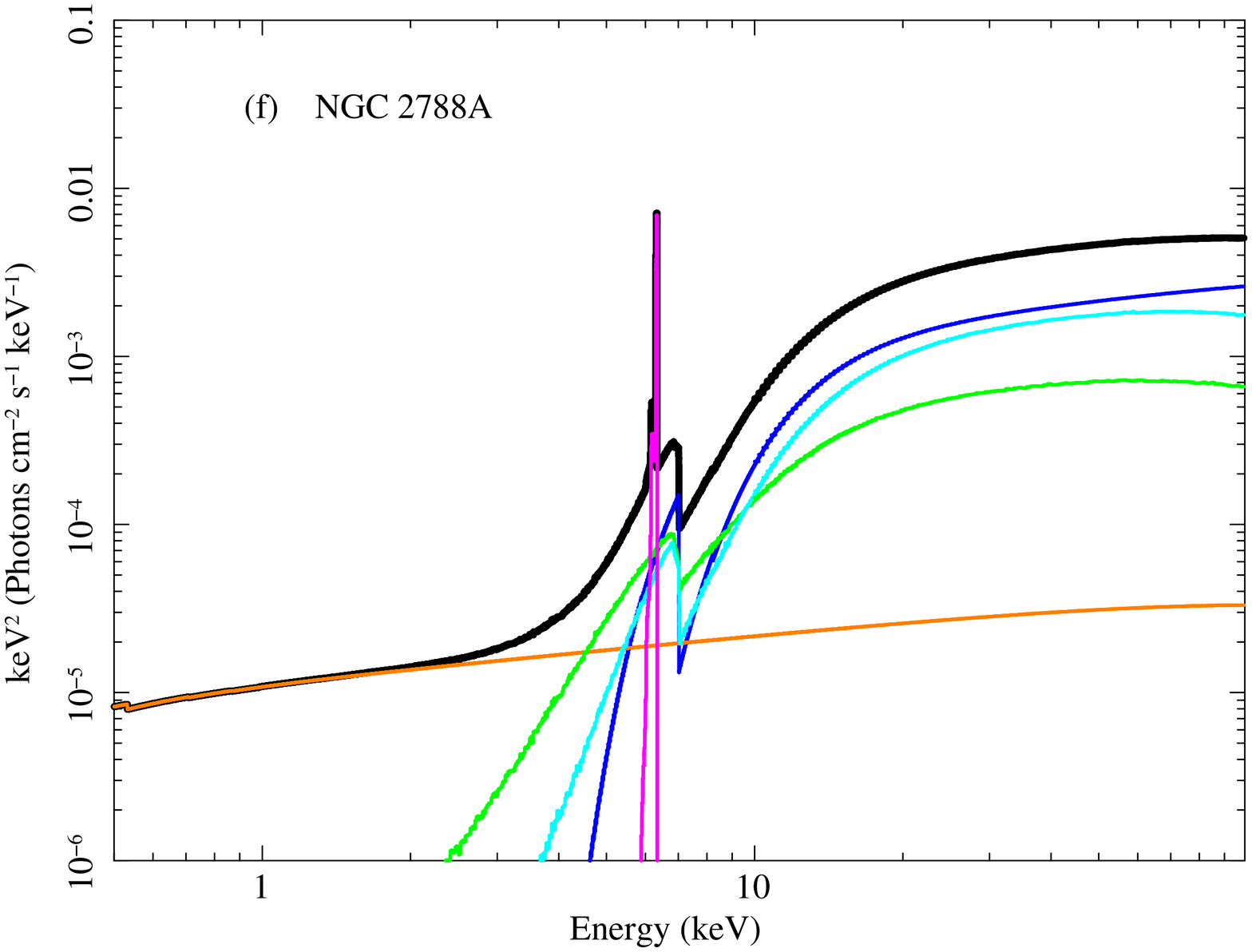}
\end{center}
\caption{
The best-fit spectral models with the MYTORUS model (left) and the Ikeda torus model (right). (a) and (d) NGC 1106. (b) and (e) UGC 03752. (c) and (f) NGC 2788A. Left: the black, blue, orange, cyan, and red lines represent the total, transmitted component, scattered component, reflection component, iron-K emission line, and emission from an optically thin thermal plasma, respectively. Right: the black, blue, orange, cyan, and red lines represent the total, absorbed reflection component, scattered component, unabsorbed reflection component, iron-K emission line, and emission from an optically thin thermal plasma, respectively.}
\end{figure*}

\section{Discussion}
\subsection{Result Summary on X-ray Spectra}
We are able to successfully fit the broadband spectra in the 0.5--100 keV band of all targets, NGC 1106, UGC 03752, and NGC 2788A with three different models, (model~1) the conventional model utilizing the \textbf{pexrav} reflection code and two numerical models based on Monte Carlo simulations, (model~2) the MYTorus model and (model~3) the Ikeda torus model. From the torus models, which are more physically self-consistent than model~1, we find that the line-of-sight column densities are larger than $10^{24}$ cm$^{-2}$ for NGC 1106 and NGC 2788A, thus identifying them as Compton-thick AGNs, while UGC 03752 is identified as a Compton-thin, heavily obscured AGN with $N_{\rm H} \sim 5.0 \times 10^{23}$ cm$^{-2}$. These identifications agree with the results by C.\ Ricci et al.\ (2016, in preperation) based on the Swift/XRT data. This fact demonstrates the effectiveness of selecting Compton-thick objects from hard X-ray surveys by using the flux ratio between above and below 10 keV and strong iron-K line emission. Table~4 lists the observed fluxes in the 0.5--2 keV, 2--10 keV, and 15--50 keV band and the intrinsic 2--10 keV luminosity based on model~3. The luminosities are in the range of $L_{\mathrm{X}} = 10^{42}-10^{43}$ erg s$^{-1}$, which are of typical Seyfert galaxies. All targets have the best-fit scattered fracton of $f_{\mathrm{scat}} > 0.5 \%$ with models~1 and 3, and hence we do not identifiy any of them as a low scattering-fraction AGN (so-called a ``new type'' AGN, \cite{Ueda07}). This is not surprizing because only $\sim$1/3 of obscured AGNs with log $N_{\rm H} >$ 23 cm$^{-2}$ are low scattering-fraction AGNs in the Swift/BAT 9-month catalog \citep{Ueda15}.

As we have mentioned earlier, model~1 should be regarded as a phenomenological model because it does not take into account the complex geometry and limited column density of the X-ray reprocessing matter. Hence, hereafter we compare the results between models~2 and 3. The difference in the MYTorus and Ikeda torus models is its geometry, in that there is more material both inside and outside of the ``doughnut'' region in the Ikeda torus for a given opening angle. Accordingly, the line-of-sight column density is generally smaller in the MYTorus model for the same equatorial column density except for the edge-on viewing case.

Model~3 gives smaller $\chi^2$ values for the same degree-of-freedom in all targets, even though the difference is only marginal ($\Delta \chi^2< 4.4$). This implies that the actual geometry of the AGN torus may be better approximated by a spherical shape rather than a doughnut shape, although we need much better quality broadband spectra to discriminate them. In fact, the two best-fit models in Figure~4 look quite similar to each other, suggesting that model degeneracy is unavoidable. There are systematic trends in the best-fit results that the column density along the equatorial plane ($N_{\mathrm{H}}^{\mathrm{Eq}}$), the inclination angle ($\theta_{\mathrm{incl}}$), and the photon index ($\Gamma$) are larger in the MYTorus case than in the Ikeda torus one, even though the 90$\%$ confidence regions of each parameter overlap between the two models (see Tables~3 and 4). This can be qualitatively understood as follows. At a given inclination angle, a larger fraction of ``unabsorbed'' reflection components from the far-side torus will be observed in the MYTorus geometry, which is constrained from the spectral curvature below 7.1 keV. As a result, a larger value of $\theta_{\mathrm{incl}}$ is obtained in the MYTorus model. Since the line-of-sight column density is a more rapidly decreasing function with decreasing inclination in the MYTorus geometry, a larger value of $N_{\mathrm{H}}^{\mathrm{Eq}}$ is required. Owning to the larger column density, stronger reflection components are predicted, which make the overall spectrum harder. To cancel this effect, the incident spectrum must become softer in the MYTorus model than in the Ikeda torus model.

The best-fit inclination angle, $\theta_{\mathrm{incl}} =$ 60.6 (MYTorus) or 60.1 (Ikeda torus), is very close to the half opening angle $\theta_{\mathrm{open}} = 60$ degree for UGC 03752. This happens because the spectrum shape below the iron-K edge (7.1 keV) seems to be dominated by an unabsorbed reflection component, which is approximated with a very hard power law rather than an absorbed continuum with a strong low energy cutoff. Similar results have been obtained from other obscured AGNs when a torus model is applied (e.g., NGC 2273, \cite{Awaki09}; NGC 3081, \cite{Eguchi11}; 3C 403, \cite{Tazaki11}). It would mean that we indeed observe the nucleus along the line-of-sight intercepting near the edge of the torus. However, considering the fact that a significant fraction of obscured AGNs show similar features, such interpretation attributing only to the geometry effect in the uniform-density tori would be unrealistic.

We interpret that these common features observed in obscured AGNs are evidence for clumpy tori (e.g., \cite{Krolik88, Nenkova08a, Nenkova08b, Kawaguchi11}), which naturally explain a large amount of unabsorbed reflection components emitted from the far-side torus. In fact, many observations imply that the tori must be clumpy; for instance, similar X-ray to mid-infrared luminosity correlations are obtained between type 1 and type 2 AGNs \citep{Gandhi09, Ichikawa12, Asmus15}, which are hard to be explained by smooth torus models. In this context, application of clumpy torus models based on Monte Carlo simulations to the broadband X-ray spectra of obscured AGNs would be required to study the detailed structure of tori, including the volume filling factor of clumps. Following the earlier works by \citet{Liu14} and \citet{Furui15}, we are working to produce numerical spectral models from clumpy tori, using the Monte Carlo simulation for Astrophysics and Cosmology (MONACO) framework \citep {Odaka11}.

\begin{table*}
\caption{Basic properties of the host galaxies}
\begin{center}
\begin{tabular}{ccccc}
\hline
\multicolumn{1}{c}{} & Parameter & NGC 1106 & UGC 03752 & NGC 2788A\\
\hline
(1)	& Galaxy type														
& S0					& S0a				& Sb							\\
(1)	& Inclination							[degree]
& 0					& 35.7				& 90.0						\\
(2)	& W1-W2								[mag]
& 0.35				& 0.90				& 0.31						\\
(2)	& W2-W3								[mag]
& 3.51 				& 3.28				& 2.75						\\
(2)	& $\log \lambda L_{\lambda}$(22 $\mu$m)	[erg s$^{-1}$]
& 43.37				& 43.59				& 42.79						\\
\hline
\multicolumn{1}{@{}l@{}}{\hbox to 0pt{\parbox{150mm}
{\footnotesize
\textbf{Notes.}\\
(1)Based on HyperLeda.\\
(2)Based on IRSA.\\
}\hss}}
\end{tabular}
\end{center}
\end{table*}

\subsection{Host Galaxy Properties}
Table~5 lists basic properties of the host galaxies of our targets taken from the HyperLeda database. Their Hubble types (S0, S0a, and Sb) are typical of normal Seyfert galaxies. The host inclination of NGC 2788A is 90 degrees (i.e., edge-on galaxy). This is consistent with the absence of an optically thin thermal component in the spectra of NGC 2788A because emission from the galactic disk may be totally absorbed by the galactic interstellar medium along the line-of-sight. It has been suggested that the flux of the soft X-ray scattered component from the nucleus may also be reduced in high inclination galaxies \citep{Honig14}. However, the two Compton-thick AGNs NGC 1106 and NGC 2788A in our sample show similar values of $f_{\mathrm{scat}}$ despite of the large difference in the host inclination. This suggests that the scattered component is not fully subject to interstellar absorption even in edge-on galaxies, most probably owning to its larger spatial extent than disk emission.

\subsection{Mid-Infrared Color and Luminosity}
\begin{figure}
\begin{center}
\FigureFile(80mm,60mm){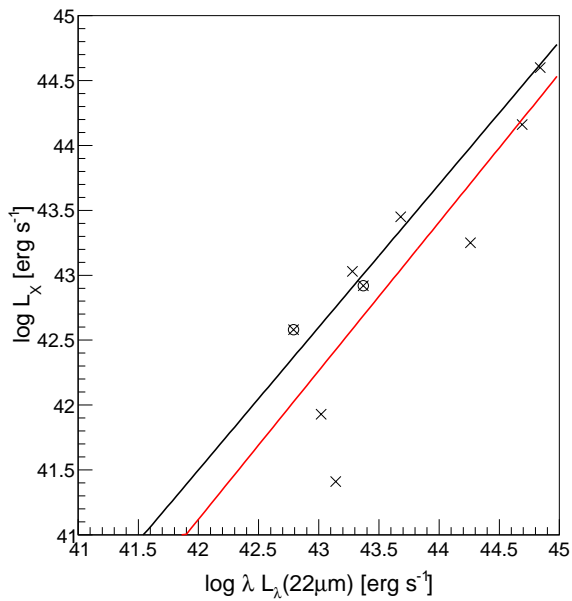}
\end{center}
\caption{
Luminosity correlation between the infrared (22 $\mu$m) and hard X-ray (14--195 keV) bands. The black line is the best linear regression line obtained from all non-blazar AGNs in the Swift/BAT 9-month catalog \citep{Ichikawa12}, while the red line is that obtained from 10 Compton-thick AGNs including  NGC 1106 and NGC 2788A (marked with circles).}
\end{figure}

We investigate the mid-infrared fluxes of our targets using the WISE catalog \citep{Wright10}, which are also summarized in Table~5. We find that there is a large variation in their WISE colors (W1-W2 vs. W2-W3) and that only one target (UGC 03752) has the colors consistent with those of ``AGN wedge'' defined by \citet{Mateos13}. This confirms the argument by \citet{Gandhi15} that the mid-infrared colors cannot be used to completely find obscured AGNs over a wide range of luminosity most probably because of host contamination.

Figure~5 plots the correlation between the WISE 22 $\mu$m luminosity ($\lambda L_{\lambda}(22 \ \mu m )$) and the intrinsic luminosity in the 2--10 keV band ($L_{\mathrm{X}}$) for a sample of nine Compton thick AGNs including NGC 1106, NGC 2788A, and those in the Swift/BAT 9-month catalog except for NGC~1365, which shows abnormally rapid variability in absorption \citep{Risaliti05}. The X-ray luminosities are taken from Table~4 for NGC 1106 and NGC 2788A, and from \citet{Ricci15} for the rest. From this sample, we obtain the linear regression form of $\log L_{\mathrm{X}} = (1.14\pm0.26)\log \lambda L_{\lambda}(22 \ \mu m) -6.96\pm11.4$ with the ordinary least-square bisector method. We also overplot the best-fit regression form obtained by \citet{Ichikawa12} for ``all'' non-blazar AGNs in the Swift/BAT 9-month catalog by converting the mid-infrared luminosities at 18 $\mu$m to those at 22 $\mu$m with equation (4) in \citet{Ichikawa12} and the X-ray luminosities in the 14--195 keV ban
 d to those in the 2--10 keV band by assuming a photon index of 1.8. As noticed, there is a trend that Compton-thick AGNs show, on average, higher mid-infrared luminosities than Compton-thin AGNs at the same hard X-ray luminosity, even though the variation is large. This agrees with the finding by \citet{Matsuta12}, who used the AKARI 9 $\mu$m and 18 $\mu$m luminosities and the ``observed'' 14--195 keV luminosities (not corrected for extinction) for all AGNs including Compton-thick ones. The averaged luminosity ratio is calculated to be $\log (\lambda L_{\lambda}(22 \ \mu m))/\log (L_{\mathrm{X}}) \sim 0.49 \pm 0.11 $ for the Compton-thick AGNs, which is larger than $\log (\lambda L_{\lambda}(22 \ \mu m))/\log (L_{\mathrm{X}}) \sim -0.02 \pm 0.05 $ derived from all non-blazar AGNs. This implies that, at least in some Compton thick AGNs, the host galaxies have larger star-forming activities that contribute to the $22 \ \mu m$ luminosities. Obviously, we need a much larger and statisti
 cally complete sample to pursue this issue further, which will be reported elsewhere.

\section{Conclusions}
Using our broadband X-ray spectra obtained with Suzaku and Swift/BAT, we confirm the heavy obscuration nature of NGC 1106, UGC 03752, and NGC 2788A, which are selected from the Swift/BAT 70-month catalog on the basis of high hardness ratio between above and below 10 keV. The conclusions of our work are summarized as follows.

\begin{itemize}
\item The broadband spectra in the 0.5--100 keV band are well reproduced with either a conventional model utilizing an analytic reflection code (\textbf{pexrav}), (2) a Monte-Carlo based torus model with a doughnut-like geometry (MYTorus), or (3) that with a nearly spherical geometry (Ikeda torus). The Ikeda torus model gives better fits than the MYTorus model with the same opening angle of the tori (60 degrees) in all targets.

\item We identify that NGC 1106 and NGC 2788A as Compton-thick AGNs and UGC 03752 as a heavily obscured Compton-thin AGN from their line-of-sight column densities.

\item In UGC 03752, these ``smooth'' torus models would require that the line-of-sight intercepts near the edge of the torus in order to explain a large amount of unabsorbed reflection components from the far-side torus. Similar results are quite commonly obtained in obscured AGNs. We interpret that this is evidence for clumpy tori.

\item We suggest that in order to reach correct interpretation of complex X-ray spectra of obscured AGNs, modeling with clumpy torus models would be necessary. This work is in progress by using the MONACO framework.
\end{itemize}

\bigskip
Part of this work was financially supported by the Grant-in-Aid for Scientific Research 26400228 (Y.U.), from the Japan Society for the Promotion of Science (JSPS) Fellows for young researchers (T.K.) and from JSPS, CONICYT-Chile ``EMBIGGEN'' Anillo (grant ACT1101), from FONDECYT 1141218 and Basal-CATA PFB--06/2007. (C.R.).

\end{document}